\documentclass{elsart}
\usepackage{latexsym}
\usepackage{amssymb}
\usepackage{amsmath}
\usepackage{rotating,graphicx}
\usepackage{epsfig}
\usepackage{psfrag}

\usepackage{caption2}

\newcommand{\beq}[0]{\begin{equation}}
\newcommand{\eeq}[0]{\end{equation}}

\begin{document}
\newlength{\caheight}
\setlength{\caheight}{12pt}
\multiply\caheight by 7
\newlength{\secondpar}
\setlength{\secondpar}{\hsize}
\divide\secondpar by 3
\newlength{\firstpar}
\setlength{\firstpar}{\secondpar}
\multiply\firstpar by 2

\parbox[0pt][\caheight][t]{\firstpar}{
  {\small \shortstack[l]{
      Corresponding author:\\
      Guido Montagna\\
      Dipartimento di Fisica Nucleare e Teorica\\
      via Bassi n.6, I-27100, Pavia\\
      e-mail: Guido.Montagna@pv.infn.it\\
      Tel: +39-0382-987742\\
      Fax: +39-0382-526938
      }}
  }
\hfill
\parbox[0pt][\caheight][t]{\secondpar}{
  \rightline
  {\shortstack[l]{
      FNT/T 2006/04}}
  }
\begin{frontmatter}
\vskip 48pt
%\title{A Non-Gaussian Approach to Value-at-Risk
%and Expected Shortfall}
\title{A Non-Gaussian Approach to Risk Measures}
\author[1,2]{Giacomo Bormetti}
\author[1,2]{Enrica Cisana}
\author[1,2]{Guido Montagna}
\author[2,1]{ and Oreste Nicrosini}
\address[1]{Dipartimento di Fisica Nucleare e Teorica, Universit\`a di Pavia\\
Via A. Bassi 6, 27100, Pavia, Italy}
\address[2]{Istituto Nazionale di Fisica Nucleare, sezione di Pavia\\
Via A. Bassi 6, 27100, Pavia, Italy}

\begin{abstract}
  Reliable calculations of financial risk require that the fat-tailed nature 
  of prices changes is included in risk measures. To this end, a non-Gaussian 
  approach to financial risk management is presented, modeling the power-law
  tails of the returns distribution in terms of a Student-$t$ distribution. Non-Gaussian
  closed-form solutions for Value-at-Risk and Expected Shortfall are obtained and
  standard formulae known in the literature under the normality assumption are
  recovered as a special case. The implications of the approach for risk management
  are demonstrated through an empirical analysis of financial time series from
  the Italian stock market and in comparison with the results of the most widely used procedures 
  of quantitative finance. Particular attention is paid to quantify the size of the errors 
  affecting the market risk measures obtained according to different methodologies, by employing a 
  bootstrap technique.
\end{abstract}


\end{frontmatter}

%%%%%%%%%%%%%%%%%%%%%%%%%%%%%%%%%%%%%%%%%%%%%%%%%%%%%%%%%%%%%%%%%%%%%%%%%%%%%%%%%%%%%%%%%%%%%%%%%%%%%%%%%%%%%%%%%%%%%%%%%%%%%%%%%%%%%

\newpage

%%%%%%%%%%%%%%%%%%%%%%%%%%%%%%%%%%%%%%%%%%%%%%%%%%%%%%%%%%%%%%%%%%%%%%%%%%%%%%%%%%%%%%%%%%%%%%%%%%%%%%%%%%%%%%%%%%%%%%%%%%%%%%%%%%%%%

\section{Introduction}
\label{s:intro}

A topic of increasing importance in modern economy and society is the 
development of reliable methods of measuring and controlling 
financial risks. According to the new capital adequacy framework, commonly
known as Basel~II accord \cite{basel}, any financial institution has to meet stringent
capital requirements in order to cover the various sources of risk that they
incur as a result of their normal operation. Basically, three different categories
of risk are of interest: credit risk, operational risk and market risk. In particular,
market risk concerns the hazard of losing money due to the fluctuations
of the prices of those instruments entering a financial portfolio and is, therefore, 
particularly important for financial risk management. 

In the financial industry today, the most widely used measure to manage 
market risk is Value-at-Risk (VaR) \cite{jorion,bouchaud_potters}. In short, VaR refers to the maximum
potential loss over a given period at a certain confidence level and can be used
to measure the risk of individual assets and portfolios of assets as well. Because
of its conceptual simplicity, VaR has become a standard component in the
methodology of academics and financial practitioners. Moreover, VaR generally provides
a reasonably accurate estimate of risk at a reasonable computational time. Still,
as discussed in the literature~\cite{bouchaud_potters,acerbietal}, VaR suffers from some inconsistencies: first, it can 
violate the sub-additivity rule for portfolio risk, which is a required property for
any consistent measure of risk, and, secondly, it doesn't quantify the typical loss
incurred when the risk threshold is exceeded. To overcome the drawbacks of
VaR, the Expected Shortfall (or Conditional VaR) is introduced, and sometimes
used in financial risk management, as a more coherent measure of risk.

Three main approaches are known in the literature and used in practice 
for calculating VaR and Expected Shortfall. The first method consists in 
assuming some probability distribution function for price changes and 
calculating the risk measures as closed-form solutions. This approach is
called parametric or analytical and is easy to implement since analytical
expressions can often be obtained. The parametric approach usually relies on
the (log)normality assumption for the returns distribution, although some 
analytical results 
using non-Gaussian functional forms are available in the literature 
\cite{finland,kamdem}. Actually, 
it is well known that empirical price returns, especially in the limit of
high frequency, do not follow the Gaussian paradigm
and are characterized by heavier tails and a higher peak than a normal 
distribution. In order to capture the leptokurtic (fat-tailed) nature of price returns, 
the historical simulation approach is often used as an alternative to the parametric method.
It employs recent historical data and risk measures are derived from the percentiles
of the distribution of real data. This method is potentially the most accurate because
it accounts for the real statistics of price changes but it is 
computationally quite demanding (especially when applied to large portfolios) and 
absolutely depending on the past history of empirical data. A third approach 
consists in Monte Carlo simulations of the stochastic dynamics of a given model
for stock price returns and in calculating risk measures according to Monte Carlo
statistics. This method, however, requires very intensive simulations to achieve 
risk measures  predictions with acceptable numerical errors. 

As a result of the present situation, reliable and possibly fast methods to calculate financial
risk are strongly demanded. Inspired by this motivation, the aim of this paper is
to present a non-Gaussian approach to market risk management and to describe its
potentials, as well as limitations, in comparison with standard procedures used
in financial analysis. To capture the excess of kurtosis of empirical data with respect
to the normal distribution, the statistics of price changes is modeled in terms of a Student-$t$ 
distribution, which is known to approximate with good accuracy the distribution 
derived from market data at a given time horizon~\cite{bouchaud_potters,mantegna_stanley} and is widely used in the financial literature. 
In the econophysics literature, the Student-$t$ distribution is 
%often referred to as a special case of 
also known as Tsallis distribution, emerging within the framework of statistical 
physics~\cite{gellmann-tsallis}. 
It has been shown in various studies~\cite{gellmann-tsallis,tsallis} that the distribution of returns can be modeled quite well
by a Tsallis distribution, which, for this reason, has been already used in a number of financial applications, ranging from option pricing~\cite{lisa} to risk analysis~\cite{mattedi_etal}. However, with respect to the investigation of Ref. \cite{mattedi_etal}, we include in our analysis the study of
the Expected Shortfall and we present, in the spirit of a parametric approach, analytical expressions 
for the risk measures in order to provide accessible results for a simple practical implementation. 
%Furthermore, in the empirical analysis of financial time series, we limit to 
%analyze a one-year data set, rather than a time series over a long time horizon, 
%in order to be close to the practice of many financial institutions.
At a variance of the recent calculation in Ref.~\cite{kamdem}, where analytical results for risk measures using Student-$t$ distributions
are presented, we critically investigate the implications of our non-Gaussian 
analytical solutions on the basis of an empirical analysis of financial data and we perform detailed
comparisons with the results of widely used procedures.

The paper is organized as follows. In Section \ref{s:closed} non-Gaussian closed-form
expressions for VaR and Expected Shortfall are derived as  generalizations of 
the analytical formulae known in the literature under the normality assumption. It is
also shown how the standard Gaussian formulae of the parametric approach are recovered, 
in the appropriate limit, as a special case. In Section \ref{s:data} an empirical analysis
of daily returns series from the Italian stock market is performed, in order to constrain the Student-$t$ parameters 
entering the formulae of Section \ref{s:closed} and to describe the ingredients needed
for the forecoming risk analysis. The latter is carried out in Section \ref{s:risk}. The implications
of the parametric non-Gaussian approach for VaR and Expected Shortfall are shown 
in Section \ref{s:risk} and compared with the results of the parametric normal method, of its
improved version known as RiskMetrics methodology and of the historical simulation. 
Particular attention is paid to quantify the size of the errors affecting the various risk
measures, by employing a bootstrap technique. Conclusions and possible 
perspectives are drawn in Section \ref{s:conclusion}.

%%%%%%%%%%%%%%%%%%%%%%%%%%%%%%%%%%%%%%%%%%%%%%%%%%%%%%%%%%%%%%%%%%%%%%%%%%%%%%%%%%%%%%%%%%%%%%%%%%%%%%%%%%%%%%%%%%%%%%%%%%%%%%%%%%%%%

%\section{Non-Gaussian closed-form expressions for Value-at-Risk and Expected Shortfall}
\section{Non-Gaussian closed-form expressions for risk measures}
\label{s:closed}

Value-at-Risk is referred to the probability of extreme losses in a portfolio value 
due to adverse market movements. In particular, for a given significance level 
$\mathcal{P}^{\star}$ (typically 1$\%$ or 5$\%$), VaR, usually denoted as $\Lambda^{\star}$, is defined as the 
maximum potential loss over a fixed time horizon $\Delta t$. In terms of price changes $\Delta S$, 
or, equivalenty, of returns $R\doteq\Delta S/S$, VaR can be computed as follows
\beq
  \label{eq:var}
  \mathcal{P}^\star \doteq \int_{-\infty}^{-\Lambda^{\star}} \mathrm{d}\Delta S~\tilde P_{\Delta t}
  (\Delta S)= S \int_{-\infty}^{-\Lambda^{\star}/ S}\mathrm{d}R~P_{\Delta t}(R),
\eeq
where $\tilde P_{\Delta t} (\Delta S)$ and $P_{\Delta t}(R)$ are the probability density functions (pdfs)
for price changes and for returns over a time horizon $\Delta t$, respectively. For financial analysts, VaR has become the standard measure used to quantify market risk because 
it has the great advantage to aggregate several risk component into a single number. 
In spite of its conceptual simplicity, VaR shows two main drawbacks: it is not 
necessary subadditive and it does not quantify the size of the 
potential loss when the threshold $\Lambda^{\star}$ is exceeded.

A quantity that does not suffer of these disadvantages is the so called Expected Shortfall (ES) or Conditional VaR (CVaR),
 $E^{\star}$. It is defined as
\beq
  \label{eq:ES}
  E^{\star} \doteq \frac{1}{\mathcal{P}^\star}\int_{-\infty}^{-\Lambda^\star}\mathrm{d}
  \Delta S~(-\Delta S)~\tilde P_{\Delta t}(\Delta S) = \frac{S}{\mathcal{P}^\star}
  \int_{-\infty}^{-\Lambda^\star/S}\mathrm{d} R~(-R)~P_{\Delta t}(R),
\eeq
with $\mathcal{P}^\star$ and $\Lambda^{\star}$ as in Eq. (\ref{eq:var}).

The standard approach in the financial literature \cite{jorion,mina_xiao} is to assume
the returns as normally distributed, with mean $m$ and variance $\sigma^2$, i.e. 
$R\sim\mathcal{N}(m,\sigma^2)$. In that case, VaR and ES analytical expressions reduce to the following
closed-form formulae
\beq
  \label{eq:var_gauss}
  \Lambda^{\star} = -mS_0 + \sigma S_0\sqrt{2}~\mathrm{erfc}^{-1}(2\mathcal{P}^\star)
\eeq
and
\beq
  \label{eq:ES_gauss}
  E^{\star} = -m S_0 + \frac{\sigma S_0}{\mathcal{P}^\star}\frac{1}
  {\sqrt{2\pi}}\exp\{-[\mathrm{erfc}^{-1}(2\mathcal{P}^\star)]^2\},
\eeq
where $S_0$ is the spot price and $\mathrm{erfc}^{-1}$ is the inverse of the complementary error function~\cite{nr}.

However, it is well known from several studies, especially in the econophysics literature~\cite{bouchaud_potters,mantegna_stanley,mandelbrot}, that the normality hypothesis 
is often inadeguate for daily returns and, more generally, for high-frequency stock price variations.
A better agreement with data is obtained using leptokurtic distributions, such as truncated L\'evy distributions %\cite{mantegna_stanley,mantegna_prl} or 
or Student-$t$ ones. Despite this interesting feature, the former family 
has the main disadvantage that it is defined only throught its characteristic function
and we have no analytic expression for the pdfs~\cite{mantegna_prl}. Moreover, in order to compute the cumulative density function (cdf),  which is a necessary 
ingredient of our analysis, we have to resort to numerical approximations.
For the reasons above, to model the returns, we make use of a Student-$t$ distribution 
defined as 
\beq
  \label{eq:student}
  \mathcal{S}^{\nu}_{m,a} (R) = \frac {1}{B(\nu/2,1/2)} \frac {a^{\nu}} {[a^{2}+(R-m)^{2}]^{\frac{\nu+1}{2}}},
\eeq
where $\nu\in (1,+\infty)$ is the tail index and $B(\nu/2,1/2)$ is the beta function. 
It is easy to verify that, for $\nu > 2$, the variance 
is given by $\sigma^{2} = a^2/(\nu - 2)$, while, for $\nu > 4$, 
the excess kurtosis reduces to $k = 6/(\nu - 4)$. Under this assumption, we obtain closed-form generalized expression for VaR and ES given by
\beq
  \label{eq:var_stud}
  \Lambda^\star = -mS_0+\sigma S_0\sqrt{\nu-2}~\sqrt{\frac{1-\lambda^\star}{\lambda^\star}}
\eeq
and
\beq
  \label{eq:ES_stud}
  E^\star = -m S_0 + \frac{\sigma S_0}{\mathcal{P}^\star B(\nu/2,1/2)}
  \frac{\sqrt{\nu-2}}{\nu-1}~[\lambda^\star]^\frac{\nu-1}{2},
\eeq
where $\lambda^\star\doteq I^{-1}_{[\nu/2,1/2]}(2\mathcal{P}^\star)$ and $I^{-1}_{[\nu/2,1/2]}$ 
\begin{figure}[t!]
  \caption{\label{fig:convergence} Convergence of the VaR (left) and ES (right) Student-$t$ formulae 
    toward Gaussian results when approaching the limit $\nu\rightarrow +\infty$.}
  \begin{minipage}[b]{0.5\textwidth}
    \begin{center}
      \psfrag{1p}[l]{\tiny 1\%}
      \psfrag{5p}[l]{\tiny 5\%}
      \psfrag{n}[rb]{\tiny Normal~~}
      \psfrag{2}[rb]{\tiny $\nu = 2.75$}
      \psfrag{3}[rb]{\tiny $\nu = 3.50$}
      \psfrag{4}[rb]{\tiny $\nu = 4.50$}
      \psfrag{1}[rb]{\tiny $\nu = 100~$}
      \psfrag{Lstar}{\footnotesize $\Lambda^\star$}
      \psfrag{Pstar}[t]{\footnotesize $\mathcal{P}^\star$} 
      \includegraphics[scale = 0.55]{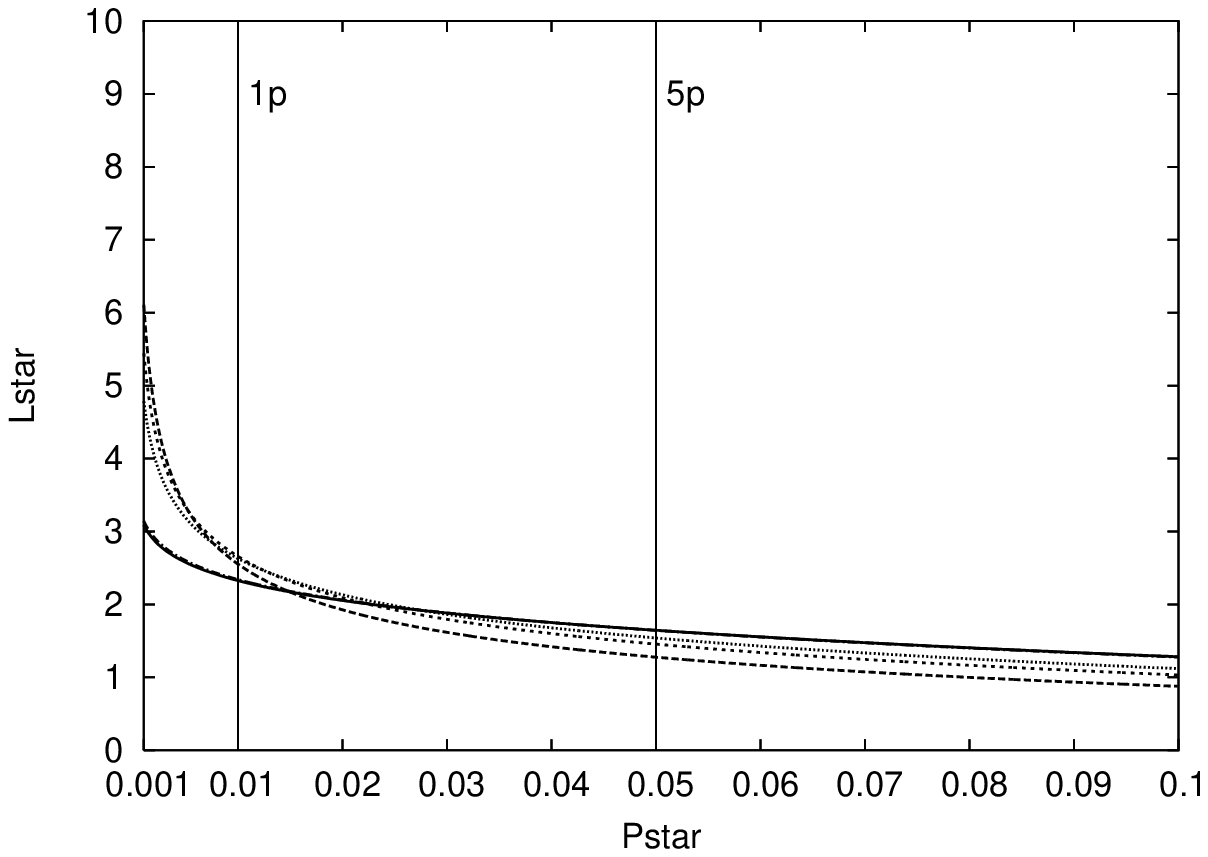}
    \end{center}
  \end{minipage}
  \begin{minipage}[b]{0.5\textwidth}
    \begin{center}
      \psfrag{1p}[l]{\tiny 1\%}
      \psfrag{5p}[l]{\tiny 5\%}
      \psfrag{n}[rb]{\tiny Normal~~}
      \psfrag{2}[rb]{\tiny $\nu = 2.75$}
      \psfrag{3}[rb]{\tiny $\nu = 3.50$}
      \psfrag{4}[rb]{\tiny $\nu = 4.50$}
      \psfrag{1}[rb]{\tiny $\nu = 100~$}
      \psfrag{Estar}{\footnotesize $\mathrm{E}^\star$}
      \psfrag{Pstar}[t]{\footnotesize $\mathcal{P}^\star$}
      \includegraphics[scale = 0.55]{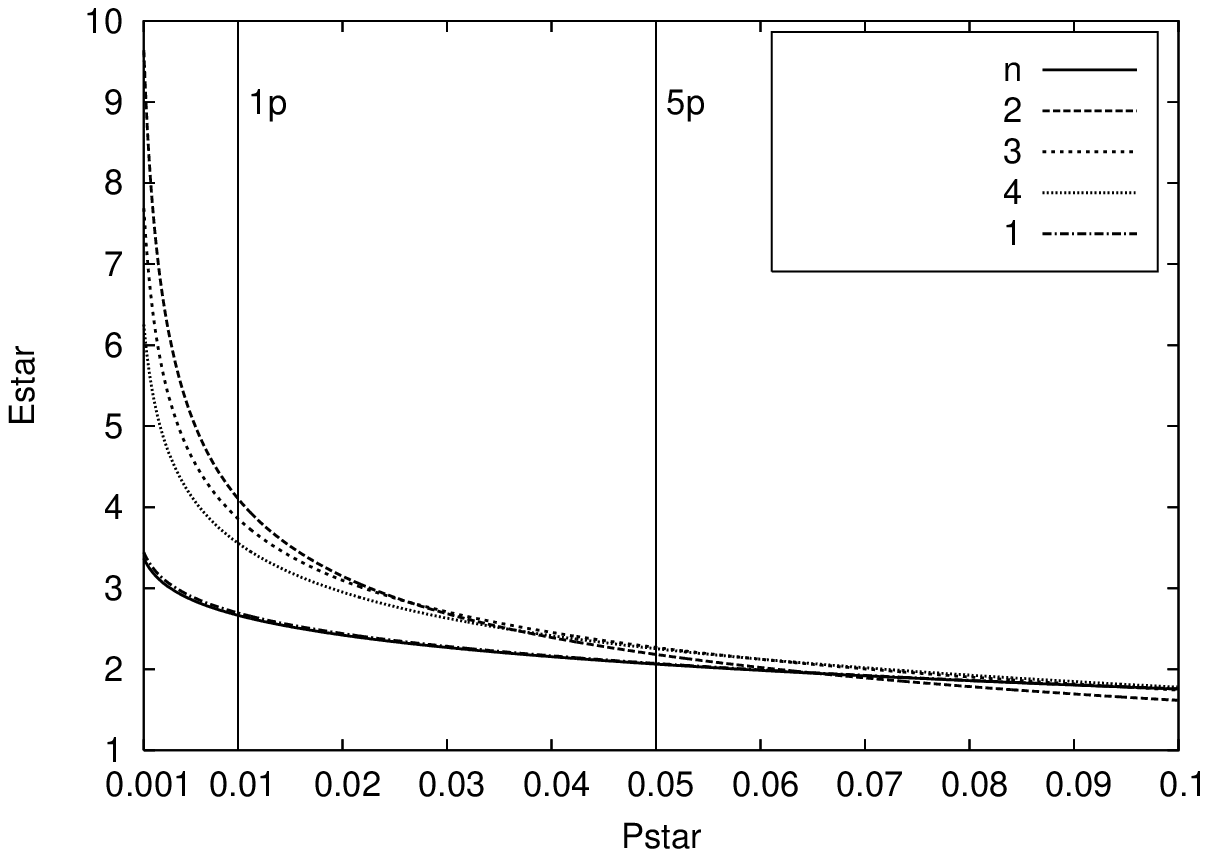}
    \end{center}
  \end{minipage}
\end{figure}
is the inverse of the incomplete beta function, according to the definition of Ref.~\cite{nr}.

\noindent As shown in Fig.~\ref{fig:convergence}, we have checked numerically the convergence of formulae~(\ref{eq:var_stud})
and (\ref{eq:ES_stud}) to the Gaussian results (\ref{eq:var_gauss}) and 
(\ref{eq:ES_gauss}),  in the appropriate limit $\nu\rightarrow +\infty$. 
We chose $\nu=2.75,3.5,4.5,100$ and $m=0$, $\sigma S_0=1$, but we checked that the value of these last parameters does not affect the convergence, as expected. As can be seen, the points corresponding to $\nu=100$ are almost coincident with the Gaussian predictions, demonstrating 
that our results correctly recover the Gaussian formulae as a special case. 
It is also worth noting that each line, corresponding to a fixed $\nu$, crosses over the Gaussian one
for a certain $\mathcal{P}^\star$. Analogously, for a fixed $\mathcal{P}^\star$,
there exists a $\nu_{\mathrm{cross}}$ value whose line crosses the 
Gaussian result at that significance level. In the light of this observation, we report  in Table~\ref{tab:cross} the values of $\nu_{\mathrm{cross}}$ corresponding to a
given $\mathcal{P}^\star$ for both VaR and ES. As can be observed, the growth of $\nu_{\mathrm{cross}}$ with $\mathcal{P}^\star$ is very rapid for VaR, 
while for ES  and for usually adopted significance values, 
$\nu_{\mathrm{cross}}$ keeps in the interval $[2.09,2.51]$. From 
this point of view, VaR and ES are quite different measures of risk, 
since the crossover values for the latter are much more stable than those associated to the first one. 
This result can be interpreted as a  consequence of ES as a more coherent risk measure than VaR.
\begin{table}[h!]
  \caption{\label{tab:cross} Values of $\nu$ crossover for VaR and ES corresponding to different
    significance levels $\mathcal{P}^\star$. }
  \begin{center}
    \begin{tabular}{@{}lccccc@{}}
      \hline
      $\mathcal{P}^\star$ & 1\% & 2\% & 3\% & 4\% & 5\% \\
      \hline 
      \hline
      $\nu_{\mathrm{cross}}$(VaR) & 2.44 & 3.21 & 5.28 & 32.38 & $\gg 100$  \\
      $\nu_{\mathrm{cross}}$(ES) & 2.09 & 2.18 & 2.28 & 2.38 & 2.51  \\
      \hline
    \end{tabular}
  \end{center}
\end{table}
%

%%%%%%%%%%%%%%%%%%%%%%%%%%%%%%%%%%%%%%%%%%%%%%%%%%%%%%%%%%%%%%%%%%%%%%%%%%%%%%%%%%%%%%%%%%%%%%%%%%%%%%%%%%%%%%%%%%%%%%%%%%%%%%%%%%%%%

\section{Empirical analysis of financial data} 
\label{s:data}

The data sets used in our analysis consist of four financial time series, composed of $N=1000$ daily returns,
from the Italian stock market. Two series are collections of data from the Italian assets Autostrade SpA and
Telecom Italia (from May 15$^{th}$ 2001 to May 5$^{th}$ 2005), while the other two correspond to the financial indexes Mib30 and Mibtel (from March 27$^{th}$ 2002 to March 13$^{th}$ 2006).
The data have been freely downloaded from Yahoo Finance \cite{yahoo}.

\begin{figure}[h!]
  \caption{\label{fig:daily} From top left clockwise: Autostrade SpA, Telecom Italia 
    (from May $15^{th}$ 2001 to May $5^{th}$ 2005), Mibtel and Mib30 
    (from March $27^{th}$ 2002 to March $13^{th}$ 2005)
    $P_>$ of negative daily returns. Points represent 
    historical complementary cdf, while dashed and solid lines correspond to Gaussian 
    and Student fits, respectively. 
    The parameters values of the fitted curves are detailed in Table~\ref{tab:param}.
  }
  \begin{minipage}[b]{0.5\textwidth}
    \begin{center}
      \psfrag{P_>}{\footnotesize $P_>$}
      \psfrag{-R}[t]{\footnotesize $~$} 
      \includegraphics[scale = 0.55]{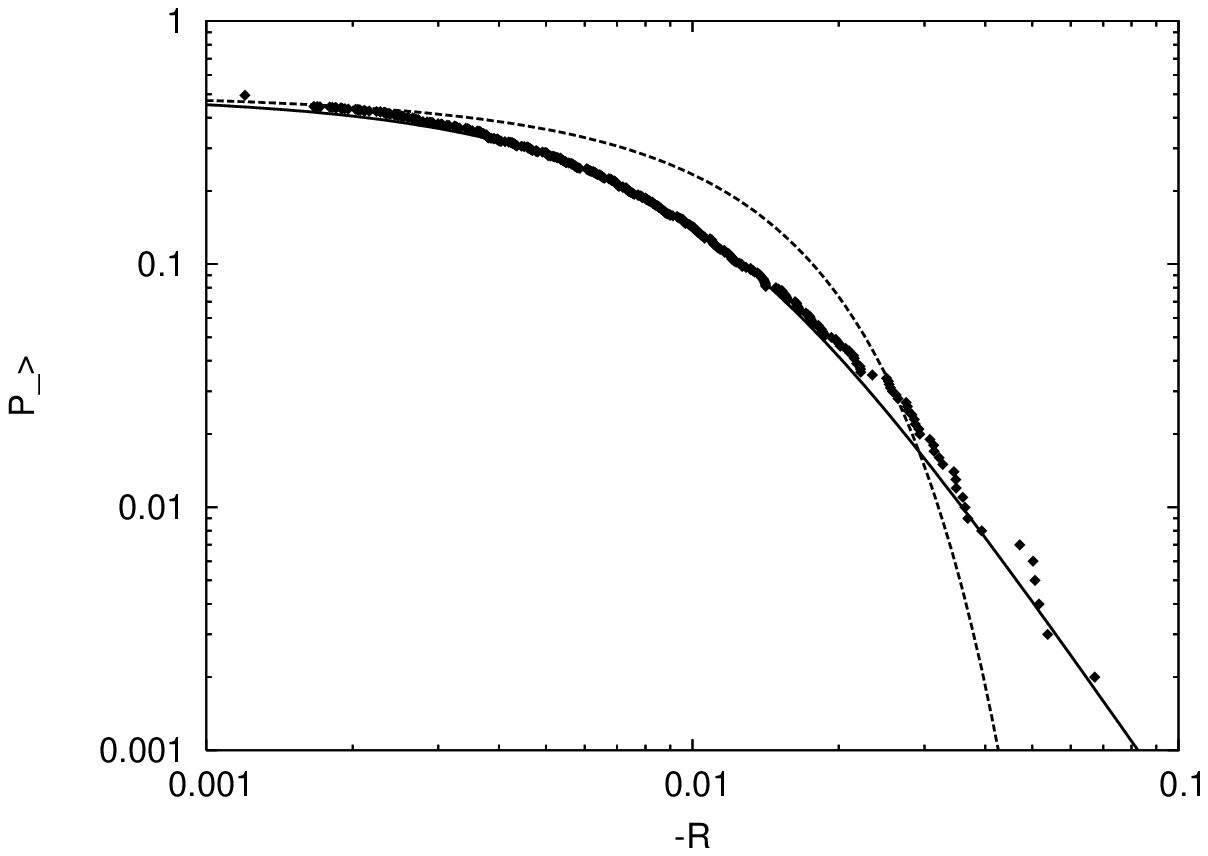}
    \end{center}
  \end{minipage}
  \begin{minipage}[b]{0.5\textwidth}
    \begin{center}
      \psfrag{A}[rb]{\tiny Data~~~}
      \psfrag{N}[rb]{\tiny Normal}
      \psfrag{S}[rb]{\tiny Student}
      \psfrag{P_>}{\footnotesize $~$}
      \psfrag{-R}[t]{\footnotesize $~$}
      \includegraphics[scale = 0.55]{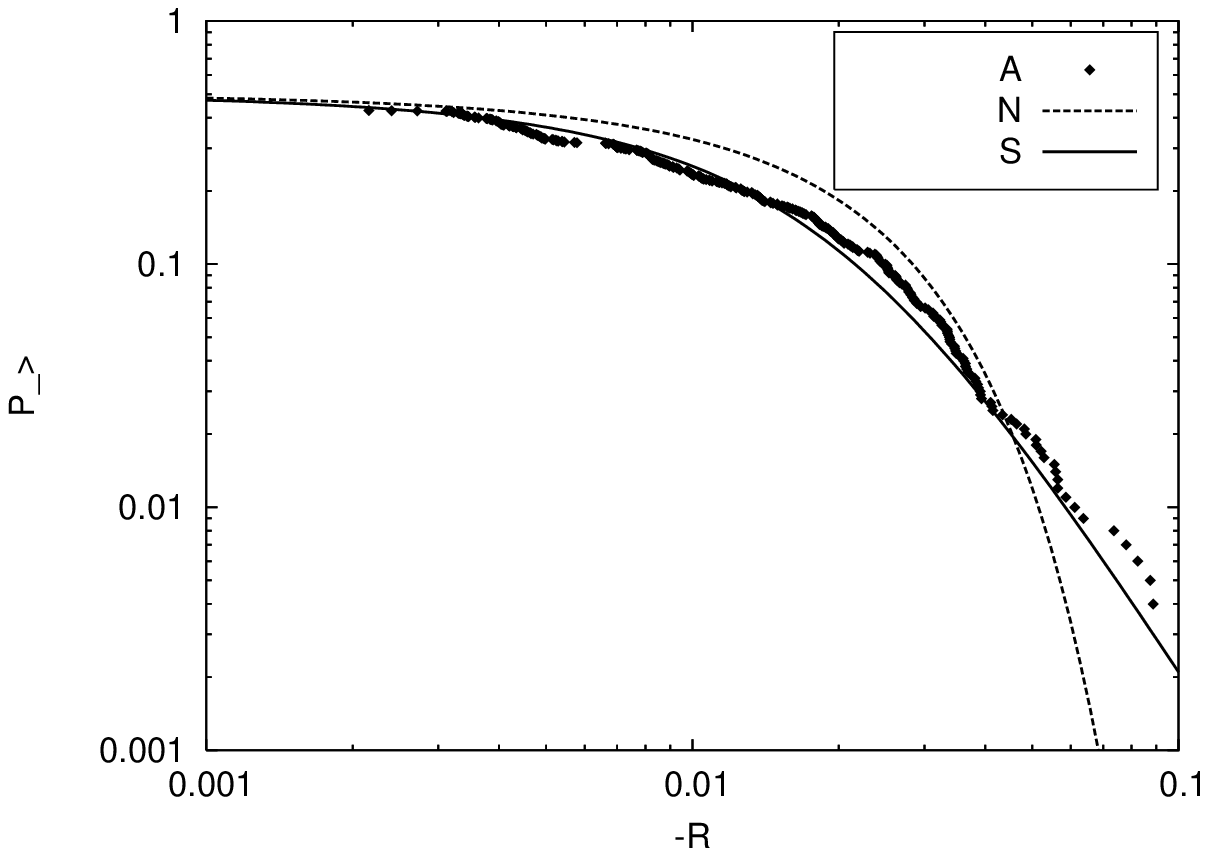}
    \end{center}
  \end{minipage}
  \begin{minipage}[b]{0.5\textwidth}
    \begin{center}
      \psfrag{P_>}{\footnotesize $P_>$}
      \psfrag{-R}[t]{\footnotesize $-R$} 
      \includegraphics[scale = 0.55]{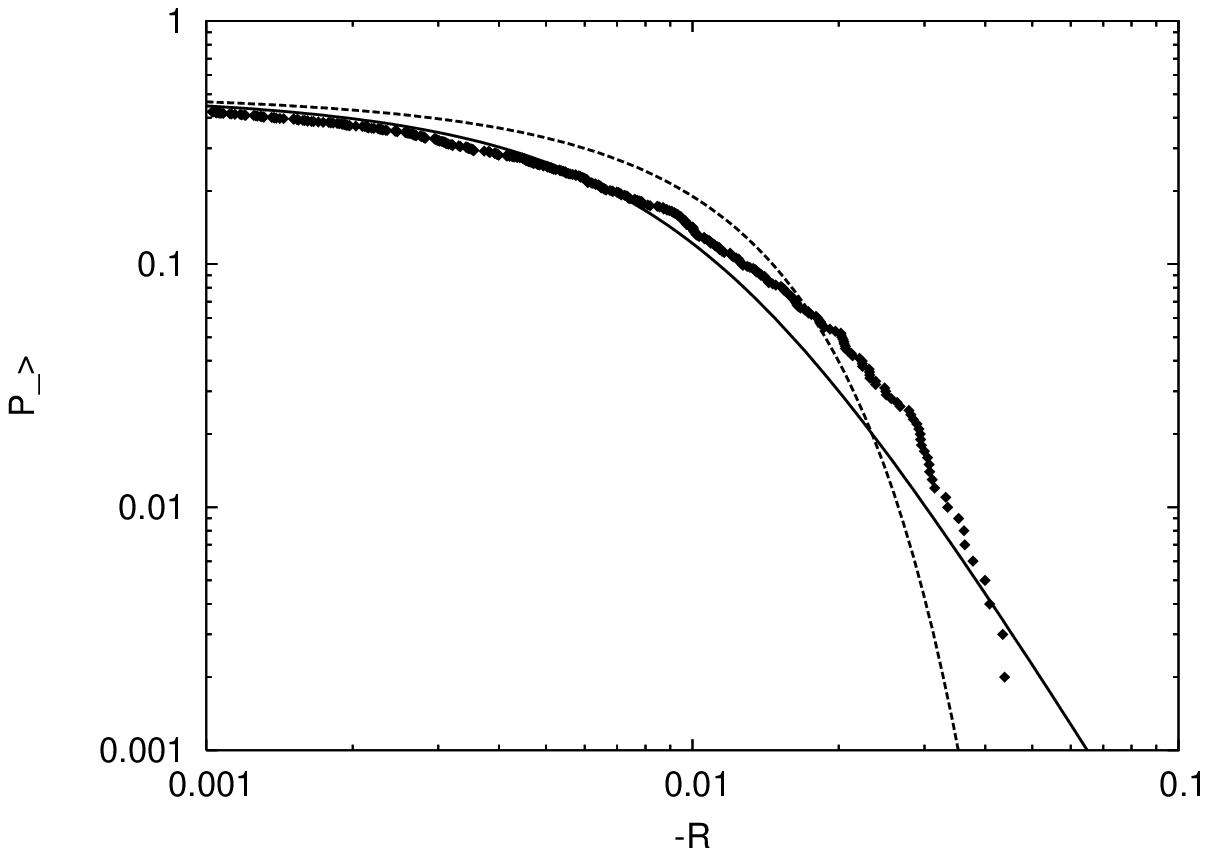}
    \end{center}
  \end{minipage}
  \begin{minipage}[b]{0.5\textwidth}
    \begin{center}
      \psfrag{P_>}{\footnotesize $~$}
      \psfrag{-R}[t]{\footnotesize $-R$}
      \includegraphics[scale = 0.55]{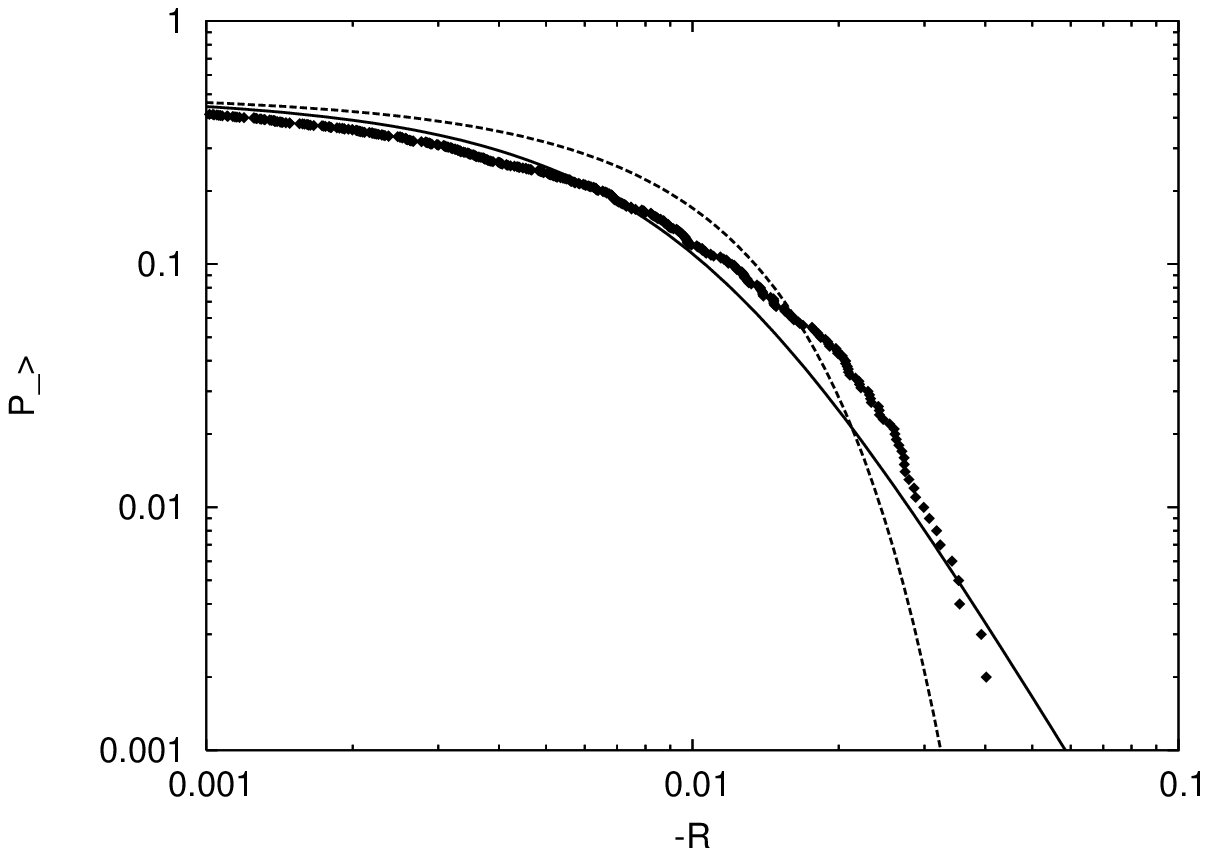}
    \end{center}
  \end{minipage}
\end{figure}

Figure~\ref{fig:daily} shows a comparison between the historical complementary cumulative density function $P_>$ of the negative
daily returns and two theoretical fits obtained using Gaussian and Student-$t$ distributions. 
The parameters values of the fitted curves, as obtained according to the likelihood
procedure described below, are displayed in Table~\ref{tab:param}.
In principle, we could perform the fit according to different methods, but we have to balance between accuracy and computational time. Therefore, we estimate mean and variance as empirical moments, 
i.e. 
\beq\label{eq:mean}
m \doteq \frac{1}{N}\sum_{i=0}^{N-1} R_{t-i}
\eeq
and
\beq\label{eq:variance}
\sigma^2 \doteq \frac{1}{N-1}\sum_{i=0}^{N-1} (R_{t-i}-m)^2,
\eeq
where $\mathbf{R}\doteq(R_{t},\ldots,R_{t-N+1})$ is 
the $N$-dimensional vector of returns. Using the above $m$ and $\sigma$ values, we derive 
a standardized vector (with zero mean and unit variance) $\mathbf{r}\doteq(r_{t},\ldots,r_{t-N+1})$, where $r_{t-i}\doteq (R_{t-i}-m)/\sigma$ for $i = 0,\ldots,N-1$. In order to find the 
best value for the tail parameter $\nu$, 
we look for the argument that minimizes the negative log-likelihood, according to the formula 
\beq\label{eq:neglog} 
\nu = \mathrm{argmin}_{~\nu>2} \left[-\sum_{i=0}^{N-1}\log \mathcal{S}^{\nu}_{0,\sqrt{\nu-2}}(r_{t-i})\right],
\eeq
where the constraint $\nu>2$ prevents the variance to be divergent and $\mathcal{S}^{\nu}_{0,\sqrt{\nu-2}}$ is as in 
Eq.~(\ref{eq:student}), with $m=0$ and $a=\sqrt{\nu-2}$. This apparently simple optimization problem
can not be solved analytically. In fact, the normalization factor in the Eq.~(\ref{eq:student}) does depend on the tail index $\nu$ in a non trivial way. Actually, the beta function $B(\nu/2,1/2)$ only admits an integral representation and therefore we implemented a numerical 
algorithm to search for the minimum.

As shown in Section \ref{s:closed}, the excess kurtosis $k$ depends only on $\nu$ and 
this provides an alternative and more efficient way to estimate the tail parameter~\cite{mattedi_etal}. 
However, this approach forces $\nu$ to be bigger than 4, while from Table~\ref{tab:param} it can
be seen that all the exponents obtained in the likelihood-based approach are smaller than 3.5.
For this reason, the implementation of the excess kurtosis method is 
inadequate for the time series under study here.
In order to test the robustness of our results, we also performed a more general three-dimensional minimization procedure over the free parameters $(m,\sigma,\nu)$. The multidimensional optimization problem was solved by using the MINUIT program from CERN library \cite{minuit}. The 
obtained numerical results are in full agreement with the previous ones, but the process is more computationally burden and more cumbersone, since it requires a lot care in avoiding
troubles related to the appearing of local minima in the minimization strategy.

%In particular, to avoid the problem of local 
%extrema, for each returns serie we performed an iterated minimization. We initialized 
%50 times the trial starting point $(m_0,\sigma_0,\nu_0)$ by randomly extracting from the
%multidimensional interval $[-0.01,0.01]\times [0.0,2.0] \times [2.01,10]$, then taking the minimum of the %local ones.

In Fig.~\ref{fig:daily} we show the cumulative distribution $P_>$ obtained 
using the empirical parameters of Table~\ref{tab:param}. As expected, we measure daily volatilities of the order of $1\%$ %(only Telecom shows a bigger one by a factor 2)
and quite negligible means ($\sim 0.01 \%$). The tail parameters fall in the range $(2.9,3.5)$,
thus confirming the strong leptokurtic nature of the returns distributions, both for 
single assets and market indexes. The quality of our fit clearly emerges from Fig.~\ref{fig:daily}, where one can see a very good agreement between Student-$t$ and historical complementary cdfs, while the Gaussian distribution fails to reproduce the data.

%Student distributions are in better agreement with historical complementary cdf than Gaussian ones.\\
Before addressing a risk analysis in the next Section, it is worth mentioning, for completeness, that 
other approaches to model with accuracy the tail exponent of the returns cdfs are discussed 
in the literature. They are based on Extreme Value Theory \cite{frey} and Hill's estimator 
\cite{epjb,clementi}. However, since they mainly focus on the tails, they require very long time series to accumulate sufficient statistics and are not considered in the present study.
\begin{table}[!]
  \caption{\label{tab:param}Mean $m$, volatility $\sigma$, and tail exponent $\nu$, 
  for Autostrade SpA, Telecom Italia, Mibtel and Mib30 time series.
  $m$ and $\sigma$ are estimated from empirical moments, while $\nu$ is obtained
  through a negative log-likelihood minimization as in Eq.~(\ref{eq:neglog}).}
  \begin{center}
    \begin{tabular}{@{}lccc@{}}
      \hline
      & $m$ & $\sigma$ & $\nu$ \\
      \hline 
      \hline
      Autostrade & $0.12\%$  & $1.38\%$  & 2.91 \\
      Telecom    & $-0.02\%$ & $2.23\%$  & 3.14 \\
      Mibtel     & $0.02\%$  & $1.03\%$  & 3.35 \\
      Mib30      & $0.02\%$  & $1.16\%$  & 3.22 \\
      \hline
    \end{tabular}
  \end{center}
\end{table}

%%%%%%%%%%%%%%%%%%%%%%%%%%%%%%%%%%%%%%%%%%%%%%%%%%%%%%%%%%%%%%%%%%%%%%%%%%%%%%%%%%%%%%%%%%%%%%%%%%%%%%%%%%%%%%%%%%%%%%%%%%%%%%%%%%%%%

\section{Risk analysis}
\label{s:risk}

In this Section we present a comparison of the results obtained estimating the market 
risk throught VaR and ES according to different methodologies. The standard approach is based on the normality assumption for the distribution of the returns.
For this case we are provided of closed-form solutions, Eqs.~(\ref{eq:var_gauss}) and (\ref{eq:ES_gauss}),
that depend on the two parameters $m$ and $\sigma$. For the time series under consideration,
the effect of the mean, as shown before, is negligible, and the surviving parameter is the volatility $\sigma$. 
Several techniques are discussed in the literature to model and forecast volatility,
based on stochastic volatilty approaches \cite{heston}, GARCH-like \cite{borland_bouchaud} and
multifractal models \cite{bacry_delour_muzy}. They usually require very long time series 
(tipically 300 high frequency returns per day over $\sim$ 5 - 10 years)
and are quite demanding from a computational point of view. 
As discussed in Section~\ref{s:data}, we limit our analysis to $1000$ daily data and we 
estimate the volatility using the empirical second moment. 
In order to avoid the problem of a uniform weight for the returns, RiskMetrics 
introduces the use of an exponential weighted moving average of 
squared returns according to the formula~\cite{mina_xiao}
\beq\label{eq:riskm}
  \sigma_{t+1\mid t}^2 \doteq \frac{1-\lambda}{1-\lambda^{N+1}}\sum_{i=0}^{N-1} \lambda^i (R_{t-i}-m)^2,
\eeq
where $\lambda \in (0,1]$ is a decay factor. 
The choice of $\lambda$ depends on the time horizon and, for $\Delta t = 1$ day, 
$\lambda = 0.94$ is the usually adopted value \cite{mina_xiao}. $\sigma_{t+1\mid t}$ represents volatility estimate at time $t$ conditional on 
the realized $\mathbf{R}$. If one considers Eq.~(\ref{eq:riskm}) as the defining equation
for an autoregressive process followed by $\sigma_{t+1\mid t}$ (coupled with $R_t = \sigma_t\epsilon_t$
with $\epsilon_t\sim\mathrm{i.i.d.}(0,1)$), Refs.~\cite{pafka_kondor,nelson} provide reasons for 
the claimed good success of the RiskMetrics methodology. In order to relax standard assumption about the return pdf without loosing the advantages coming from
a closed-form expression, we presented in Section~\ref{s:closed} generalized 
formulae for VaR and ES based on a Student-$t$ modeling of price returns. In this framework, the tail index $\nu$ emerges  as a third relevant parameter, which is possible to 
constrain using a maximum likelihood technique, as previously described. As a benchmark of all our results, we also quote VaR and ES estimates following a historical approach, 
which is a procedure widely used in the practice. According to this
approach, after ordering the $N$ data in increasing order, we consider the $[N\mathcal{P^\star}]^\mathrm{th}$
return $R_{([N\mathcal{P^\star}])}$ as an estimate for VaR 
and the empirical mean over first $[N\mathcal{P^\star}]$ returns as an estimate for ES
\footnote{The symbol $[~]$ stands for integer part, while $R_{(j)}$ is the standard notation for the 
$j^{th}$ term of the order statistic of $\mathbf{R}$. Since $N \gg 1$ we neglect the fact
that the $p^{th}$ entry is a biased estimator of the $p/N$-quantile, i.e. $\mathbb{E}[R_{(p)}]=p/(N+1)$.}.

\begin{figure}[t!]
  \caption{\label{fig:es_var}
    VaR $\Lambda^\star$ (upper panel) and ES $\mathrm{E}^\star$ (lower panel) central values with 68\% CL intervals for Autostrade SpA (left) and 
    for Mib30 (right), according to the four different methodologies discussed in the text. 
    The significance level $\mathcal{P}^\star$ is fixed to $1\%$ (circles, solid lines) and $5\%$ (triangles, dashed lines). 
  }
  \begin{minipage}[b]{0.5\textwidth}
    \begin{center}
      \psfrag{lstar}[c]{\footnotesize $\Lambda^\star$}
      \psfrag{ML}[c]{\tiny $\mathrm{Student-}t$} 
      \psfrag{N}[c]{\tiny $\mathrm{Normal}$} 
      \psfrag{H}[c]{\tiny $\mathrm{Historical}$}
      \psfrag{EW}[c]{\tiny $\mathrm{~~~~~RiskMetrics}$}
      \includegraphics[scale = 0.55]{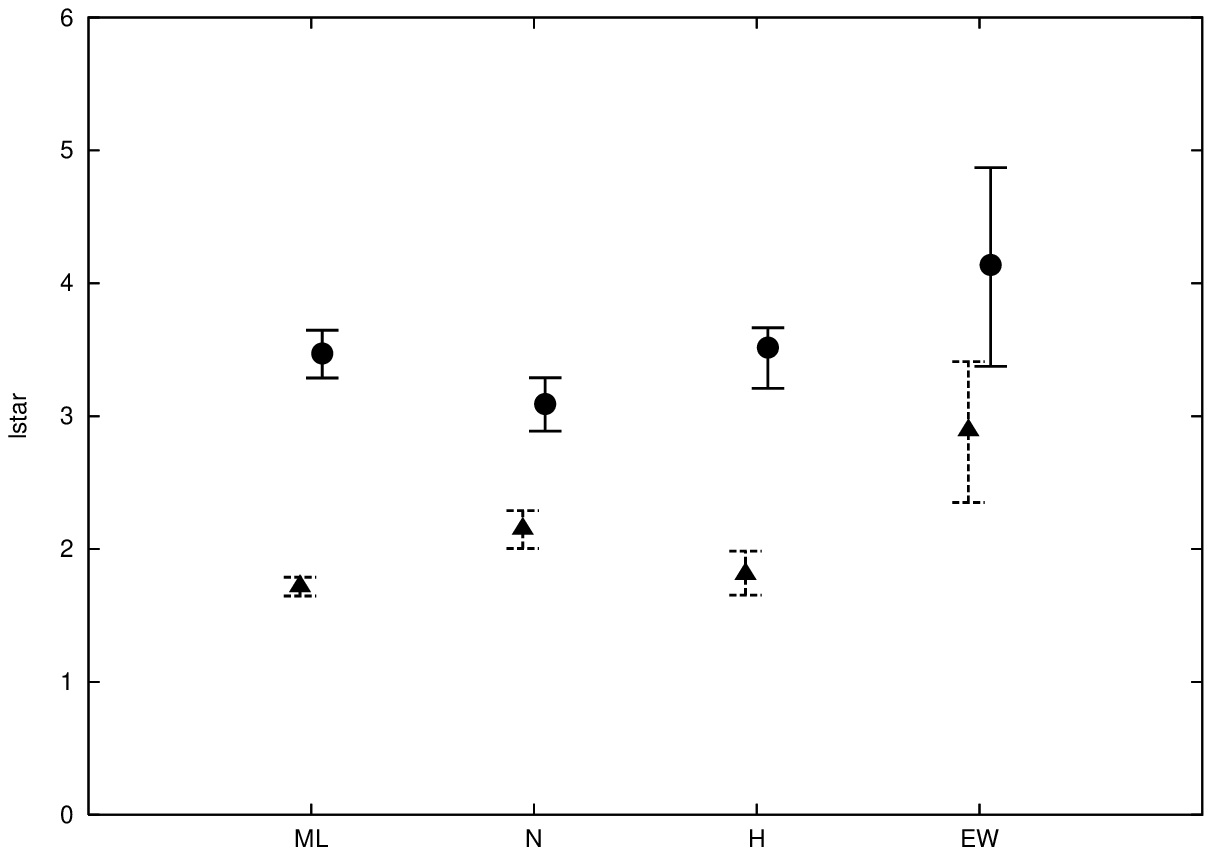}
    \end{center}
  \end{minipage}
  \begin{minipage}[b]{0.5\textwidth}
    \begin{center}
      \psfrag{lstar}[c]{\footnotesize $~$}
      \psfrag{ML}[c]{\tiny $\mathrm{Student-}t$} 
      \psfrag{N}[c]{\tiny $\mathrm{Normal}$} 
      \psfrag{H}[c]{\tiny $\mathrm{Historical}$}
      \psfrag{EW}[c]{\tiny $\mathrm{~~~~~RiskMetrics}$} 
      \psfrag{1}[rb]{\tiny $\mathcal{P}^\star 1\%$}
      \psfrag{5}[rb]{\tiny $\mathcal{P}^\star 5\%$}
      \includegraphics[scale = 0.55]{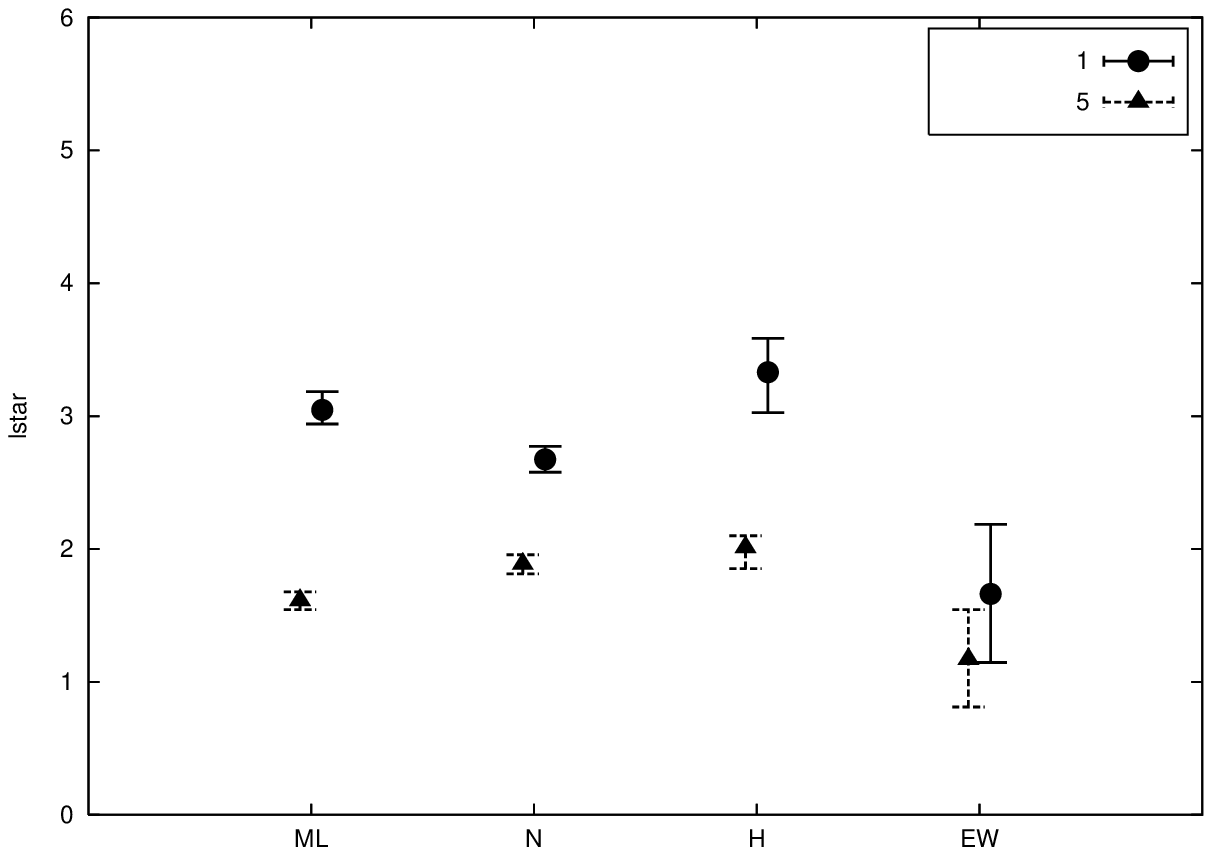}
    \end{center}
  \end{minipage}
  \begin{minipage}[b]{0.5\textwidth}
    \begin{center}
      \psfrag{estar}[c]{\footnotesize $\mathrm{E}^\star$}
      \psfrag{ML}[c]{\tiny $\mathrm{Student-}t$} 
      \psfrag{N}[c]{\tiny $\mathrm{Normal}$} 
      \psfrag{H}[c]{\tiny $\mathrm{Historical}$}
      \psfrag{EW}[c]{\tiny $\mathrm{~~~~~RiskMetrics}$}
      \psfrag{AUTS}[ct]{\footnotesize $\mathrm{Autostrade}$}
      \includegraphics[scale = 0.55]{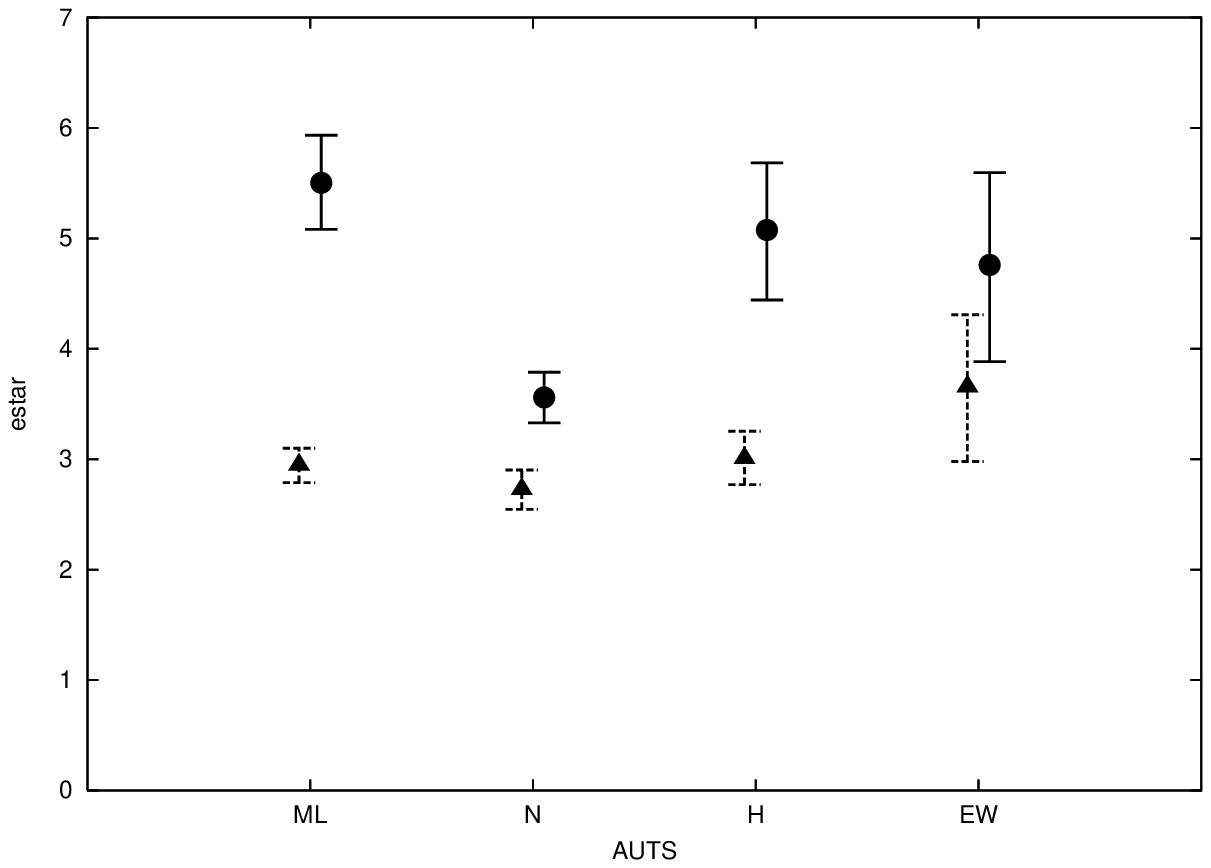}
    \end{center}
  \end{minipage}
  \begin{minipage}[b]{0.5\textwidth}
    \begin{center}
      \psfrag{estar}[c]{\footnotesize $~$}
      \psfrag{ML}[c]{\tiny $\mathrm{Student-}t$} 
      \psfrag{N}[c]{\tiny $\mathrm{Normal}$} 
      \psfrag{H}[c]{\tiny $\mathrm{Historical}$}
      \psfrag{EW}[c]{\tiny $\mathrm{~~~~~RiskMetrics}$}
      \psfrag{MIB3}[ct]{\footnotesize $\mathrm{Mib30}$}
      \includegraphics[scale = 0.55]{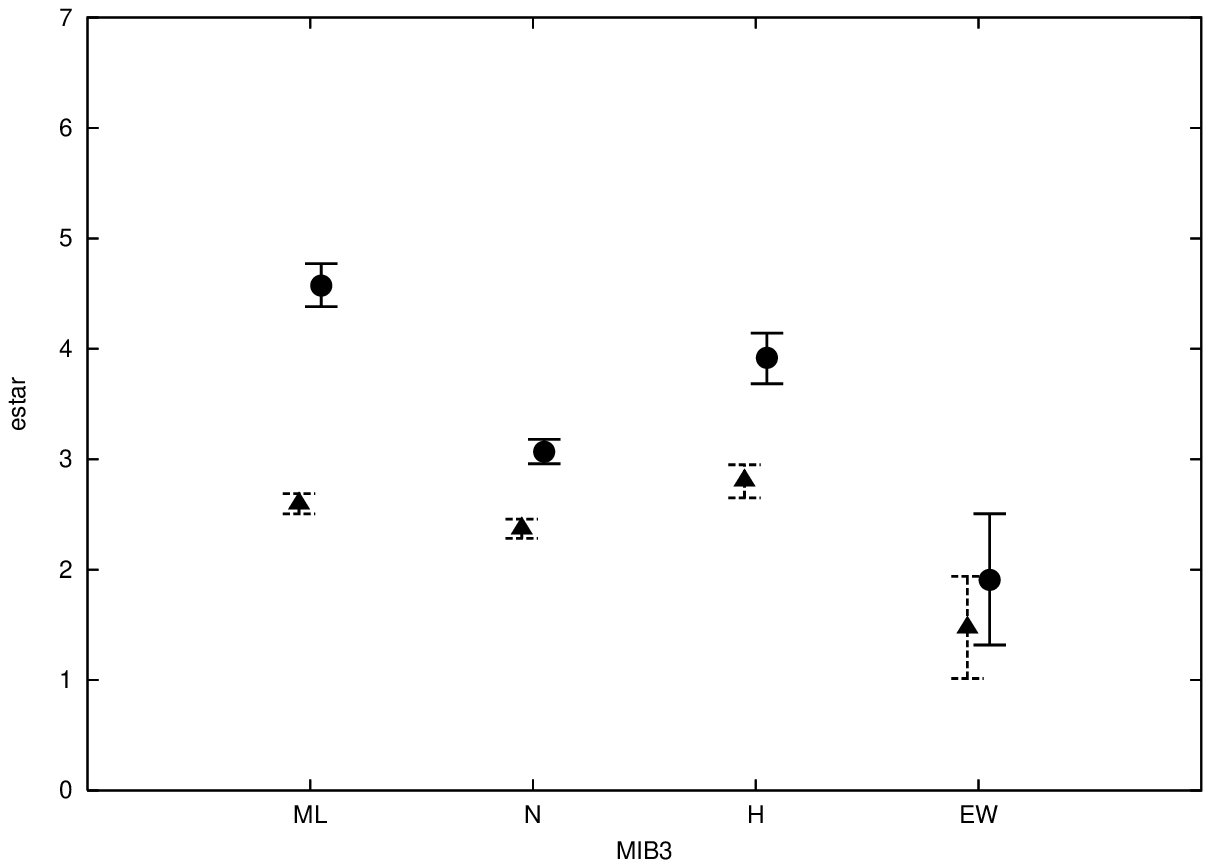}
    \end{center}
  \end{minipage}
\end{figure}

At a variance with respect to previous investigations \cite{mattedi_etal,pafka_kondor}, 
we also provide 68\% confidence level (CL) intervals associated to the parameters. 
In this way we can estimate VaR and ES dispersion.
To this extent, we implement a bootstrap technique \cite{efron_tibshirani}. 
Given the $N$ measured returns, we generate $M=1000$ sinthetic copies of $\mathbf{R}$,
$\{\mathbf{R}^*_j\}$, with $j=1,\ldots,M$, by random sampling with replacement according to the probability 
$p=(1/N,\ldots,1/N)$. For each $\mathbf{R}^*_j$ we estimate the quantities of interest and we obtain 
bootstrap central values and confidence levels. For example, we use for the mean the relations
\beq\label{eq:mean_boot}
  m^*_b\doteq\frac{1}{M}\sum_{j=1}^M m^*_j
  \quad\mathrm{with}\quad
  m^*_j=\frac{1}{N}\sum_{i=0}^{N-1}(R^*_j)_{t-i}
\eeq
and we define the $1-2\alpha$ CL interval as $[m_\alpha^*,m_{1-\alpha}^*]$, with $m_a^*$ such that 
$P(m^*\leq m^*_a)=a$ and $a=\alpha,1-\alpha$. For 68\% CL, $\alpha=16$\%. In 
Fig.~\ref{fig:es_var} and Tables~\ref{tab:boot}, \ref{tab:var}, \ref{tab:es}
we quote results according to $m-(m_b^*-m_\alpha^*)+(m_{1-\alpha}^*-m_b^*)$. In this way, we use the bootstrap approach in order to estimate the dispersion of the mean around the measured value, $m$. Analogously for all
\begin{table}[!]
  \caption{\label{tab:boot}Parameters values and bootstrap estimates for the 68\% CL intervals
    for the time series as in Table~\ref{tab:param}.
  }
  \begin{center}
    \begin{tabular}{@{}lccccc@{}}
      \hline
      & $m$ & $\sigma$ & $\sigma_{t+1\mid t}$ & $\nu$ & $R_{(10)}$ \\
      \hline 
      \hline
      Autostrade & $0.12^{+0.04}_{-0.05}\%$  & $1.38^{+0.08}_{-0.10}\%$  & $1.83^{+0.31}_{-0.33}\% $ & $2.91^{+0.20}_{-0.21}$ & $-3.51_{-0.15}^{+0.31}\%$\\ 
      Telecom    & $-0.02^{+0.06}_{-0.07}\%$ & $2.23^{+0.11}_{-0.11}\%$  & $1.54^{+0.42}_{-0.47}\% $ & $3.14^{+0.21}_{-0.22}$ & $-6.14_{-1.35}^{+0.87}\%$\\ 
      Mibtel     & $0.02^{+0.02}_{-0.04}\%$  & $1.03^{+0.03}_{-0.04}\%$  & $0.69^{+0.19}_{-0.20}\%$ & $3.35^{+0.18}_{-0.19}$ & $-2.96^{+0.25}_{-0.24}\%$ \\  
      Mib30      & $0.02^{+0.03}_{-0.04}\%$  & $1.16^{+0.03}_{-0.05}\%$  & $0.72^{+0.22}_{-0.22}\%$ & $3.22^{+0.15}_{-0.16}$ & $-3.33^{+0.30}_{-0.25}\%$ \\  
      \hline
    \end{tabular}
  \end{center}
\end{table}
the other parameters.

Table~\ref{tab:boot} shows central values and estimated 68\% CL intervals for the daily 
returns series under study. These numerical results come from a straightforward application of the resampling technique. 
It is worth mentioning that it is possible, and sometimes necessary, to use improved versions of the bootstrap. 
As a rule of thumb, we consider the boostrap approach accurate when, given a generic parameter, the difference between its empirical value and the boostrap central value estimate is close to zero and 68\% CL interval is symmetric to a good approximation. In our numerical simulation,
we measured a systematic non zero bias for $\sigma_{t+1\mid t}$ and from Table~\ref{tab:boot} it is quite evident the asymmetry of $R_{([N\mathcal{P^\star}])}$ intervals for both Autostrade and Telecom data.
We can, therefore, consider the corresponding CL intervals as a first approximation of the right ones, since bias and skewness corrections would require sofisticated and ad-hoc  techniques~\cite{efron_tibshirani},
which are beyond the scope of the present work.

In Fig.~\ref{fig:es_var} we show VaR and ES central values and 68\% CL bars 
for Autostrade SpA and Mib30, corresponding to 1\% and 5\% significance level and according 
to the four methodologies previously described. In Tables~\ref{tab:var} and \ref{tab:es} we detail all the numerical results, including also Telecom Italia and Mibtel data. As already noted in Ref.~\cite{pafka_kondor}, at 5\% significance level
Student-$t$ and Normal approaches are substantially equivalent, but here such a statement
sounds more statistically robust, thanks to the bootstrap 68\% confidence levels and to the 
comparison with the historical simulation. At this significance level, we register for VaR a different behaviour between single assets and indexes.
While assets show the best agreement between the Student-$t$ and historical approaches 
(see also Table~\ref{tab:var}), for Mib30 and Mibtel data we observe the best agreement between the Normal and historical methodology. In order to enforce this empirical evidence, it would be necessary to analyze additional time series to see to what extent this difference 
between assets and indexes holds. From 
Fig.~\ref{fig:es_var}, Table~\ref{tab:var} and Table~\ref{tab:es} it can also be seen that $\Lambda^\star$ and $\mathrm{E}^\star$ central values calculated according to RiskMetrics   
methodology are quite fluctuating and characterized by the largest CL bars.
The decreasing of $\mathcal{P}^\star$ traduces in a major differentiation of the different approaches. 
In general, we obtain the best agreement between 
the Student-$t$ approach and the historical simulation, both for $\Lambda^\star$ and 
$\mathrm{E}^\star$, whereas, as before, the RiskMetrics methodology overestimates or
underestimates the results of the historical evaluation and is affected by rather large 
uncertainties. 

To conclude, we would like to note that we expect, from the results shown in Fig.~\ref{fig:convergence} and Table~\ref{tab:cross}, that, for a fixed significance level, there exists a crossover value, $\nu_{cross}$,
below which the generalized Student-$t$ VaR and ES formulae underestimate the Gaussian 
predictions. This effect was already mentioned in Ref.~\cite{mattedi_etal}, but the analytical formulae here derived allow us to better characterize it. Under the hypothesis of a Student-$t$ distribution, the crossover value does not depend on the first and second moments and, therefore, the knowledge,
for a given time series, of the tail exponent only is sufficient to conclude, a priori, whether 
the fat-tailed results for VaR and ES will underestimate or not the 
corresponding Gaussian estimates.

\begin{table}
  \begin{center}
    \caption{\label{tab:var}Estimated VaR values (mean and 68\% CL interval) for 
    1$\%$ and 5$\%$ significance levels 
      from Autostrade SpA, Telecom Italia, Mib30 and Mibtel. For each time series,  
      the results of Student-$t$ and Normal fit, historical simulation and RiskMetrics methodology
      are shown.
    }
  \end{center}
  \begin{center}
    \begin{tabular}{ll|l|l|l|l}
      \hline
      & & Student-$t$ & Normal & Historical & RiskMetrics\\
      \hline
      \hline
      Autostrade & VaR 1\% & $3.472_{-0.185}^{+0.175}$ & $3.091_{-0.204}^{+0.197}$ & $3.516_{-0.306}^{+0.149}$ & $4.138_{-0.764}^{+0.733}$\\
      & VaR 5$\%$ &  $1.717_{-0.071}^{+0.071}$ & $2.150_{-0.145}^{+0.139}$ & $1.810_{-0.156}^{+0.175}$ & $2.890_{-0.540}^{+0.520}$\\
      \hline
%      \hline
      Telecom & VaR 1\% & $5.900_{-0.230}^{+0.279}$ & $5.200_{-0.277}^{+0.275}$ & $6.137_{-0.866}^{+1.348}$ & $3.595_{-1.085}^{+0.990}$\\
      & VaR 5$\%$ &  $3.121_{-0.137}^{+0.135}$ & $3.682_{-0.202}^{+0.214}$ & $3.398_{-0.127}^{+0.110}$ & $2.548_{-0.777}^{+0.694}$\\
      \hline
%      \hline
      Mib30 & VaR 1\% & $3.047_{-0.105}^{+0.106}$ & $2.675_{-0.096}^{+0.097}$ & $3.331_{-0.304}^{+0.255}$ & $1.662_{-0.516}^{+0.524}$\\
      & VaR 5$\%$ & $1.612_{-0.067}^{+0.066}$ & $1.885_{-0.072}^{+0.073}$ & $2.010_{-0.157}^{+0.090}$ & $1.169_{-0.358}^{+0.375}$\\
      \hline
%      \hline
      Mibtel & VaR 1\% & $2.718_{-0.092}^{+0.097}$ & $2.378_{-0.084}^{+0.088}$ & $2.967_{-0.255}^{+0.240}$ & $1.581_{-0.449}^{+0.453}$\\
      & VaR 5$\%$ &  $1.454_{-0.062}^{+0.062}$ & $1.674_{-0.065}^{+0.065}$ & $1.811_{-0.173}^{+0.150}$ & $1.110_{-0.316}^{+0.324}$\\
      \hline
    \end{tabular}
  \end{center}
\end{table}
\begin{table}
  \begin{center}
    \caption{\label{tab:es}Estimated ES values (mean and 68\% CL interval) for 1\% and 5\%
significance levels. Time series and methodologies as in Table~\ref{tab:var}. 
    }
  \end{center}
  \begin{center}
    \begin{tabular}{ll|l|l|l|l}
      \hline
      & & Student-$t$ & Normal & Historical & RiskMetrics\\
      \hline
      \hline
      Autostrade & ES 1\% & $5.503_{-0.421}^{+0.431}$ & $3.559_{-0.231}^{+0.229}$ & $5.076_{-0.634}^{+0.607}$ & $4.759_{-0.876}^{+0.837}$\\
      & ES 5\% & $2.946_{-0.159}^{+0.153}$ & $2.727_{-0.182}^{+0.175}$ & $3.006_{-0.235}^{+0.248}$ & $3.655_{-0.677}^{+0.653}$\\
      \hline
      Telecom & ES 1\% & $8.912_{-0.583}^{+0.579}$ &  $5.954_{-0.310}^{+0.311}$ & $9.685_{-1.475}^{+1.456}$ & $4.116_{-1.250}^{+1.133}$\\
      & ES 5\% & $5.035_{-0.246}^{+0.242}$ & $4.613_{-0.246}^{+0.248}$ & $5.320_{-0.466}^{+0.478}$ & $3.190_{-0.969}^{+0.879}$\\
      \hline
      Mib30 & ES 1\% & $4.572_{-0.191}^{+0.199}$ & $3.068_{-0.109}^{+0.111}$ & $3.918_{-0.234}^{+0.223}$ & $1.908_{-0.590}^{+0.599}$\\
      & ES 5\% & $2.596_{-0.091}^{+0.093}$ & $2.369_{-0.086}^{+0.088}$ & $2.804_{-0.155}^{+0.145}$ & $1.471_{-0.458}^{+0.467}$\\
      \hline
      Mibtel & ES 1\% & $4.021_{-0.171}^{+0.179}$ & $2.728_{-0.094}^{+0.099}$ & $3.501_{-0.224}^{+0.215}$ & $1.815_{-0.516}^{+0.524}$\\
      & ES 5\% & $2.314_{-0.081}^{+0.084}$ & $2.106_{-0.077}^{+0.078}$ & $2.524_{-0.136}^{+0.128}$ & $1.399_{-0.400}^{+0.399}$\\
      \hline
    \end{tabular}
  \end{center}
\end{table}

%%%%%%%%%%%%%%%%%%%%%%%%%%%%%%%%%%%%%%%%%%%%%%%%%%%%%%%%%%%%%%%%%%%%%%%%%%%%%%%%%%%%%%%%%%%%%%%%%%%%%%%%%%%%%%%%%%%%%%%%%%%%%%%%%%%%%

\section{Conclusions}
\label{s:conclusion}

In this paper we have presented a careful analysis of financial market risk measures 
in terms of a non-Gaussian (Student) model for price fluctuations. We have
derived closed-form parametric formulae for Value at Risk and Expected Shortfall that generalize standard 
expressions known in the literature under the normality assumption and can be used to
obtain reliable estimates for the risk associated to a single asset or a portfolio of assets. 
The obtained non-Gaussian parametric formulae have been shown to be able to
capture accurately the fat-tailed nature of financial data and, when specified in terms of the
model parameters optimized by means of an empirical analysis of real daily returns series, 
have been found to be in good agreement with a full historical evaluation. Moreover,
the risk measures obtained through our model show non negligible differences
with respect to the widely used Normal and RiskMetrics methodologies, indicating that the 
approach may have helpful implications for practical applications in the field of financial
risk management. We also proposed a bootstrap-based technique to estimate the
size of the errors affecting the risk measures derived through the different procedures, 
in order to give a sound statistical meaning to our comparative analysis. 

As far as possible perspectives are concerned, it would be interesting to investigate
to what extent our conclusions, drawn from an analysis of a sample of Italian financial
data, apply also to other financial markets. In particular, one could check whether, at 
a given significance level, statistically relevant differences are present between the results valid for
a single asset and those relative to a portfolio of assets, as our analysis seems to indicate, 
at least for 5\% VaR. Another interesting development concerns the comparison between the predicitions for VaR and 
 ES of our model with  the corresponding ones derived by means of other statistical procedures
to measure tail exponents known in the literature~\cite{frey,epjb,clementi}, as well as with the results from simulations of advanced models of the financial market dynamics, such as GARCH-like and multifractal models~\cite{borland_bouchaud,bacry_delour_muzy}.

\vskip 8pt\noindent
{\bf Acknowledgments} \\ 
We are grateful to Enrico Melchioni of FMR Consulting for continuous interest in our work and
very useful discussion about financial risk. We wish to thank Paolo Pedroni and Alberto Rotondi
for informative discussions about the bootstrap, as well as for suggestions about the
relative bibliography. We acknowledge precious collaboration with Andrea Fontana
concerning software issues and, in particular, the CERN tool quoted in Ref. \cite{minuit}.


\begin{thebibliography}{99}

\bibitem{basel} Basel Committee on Banking Supervision, 
  International Convergence 
  of Capital Measurement and Capital Standards: a Revised Framework, 2004 
  [\verb|http://www.bis.org/publ/bcbs107.htm|].

\bibitem{jorion} P. Jorion, Value at Risk: the New Benchmark for Managing Financial Risk, McGraw Hill, 2001. 
  
\bibitem{bouchaud_potters} J.P. Bouchaud and M. Potters, Theory of Financial Risk 
  and Derivative Pricing: from Statistical Physics to Risk Management, Cambridge University Press, Cambridge, 2003.
  
\bibitem{acerbietal} See, for example, C. Acerbi, C. Nordio and C. Sirtori, 
  Expected Shortfall as a tool for financial risk management, cond-mat/0102304 and
  references therein.

\bibitem{finland} V.P. Heikkinen and A. Kanto, Journal of Risk 4 (2002) 77.
  
\bibitem{kamdem} J.S. Kamdem, Int. J. Theoretical and Applied Finance 8 (2005) 537. 
  
\bibitem{mantegna_stanley} R.N. Mantegna and H.E. Stanley, An Introduction to Econophysics: Correlations 
  and Complexity in Finance, Cambridge University Press, Cambridge, 2000.

\bibitem{gellmann-tsallis} M. Gell-Mann and C. Tsallis, Nonextensive Entropy - 
  Interdisciplinary Applications, Oxford University Press, New York, 2004.

\bibitem{tsallis} C. Tsallis, C. Anteneodo, L. Borland and R. Osorio, Physica A 324 (2003) 89.

\bibitem{lisa} L. Borland, Quantitative Finance 2 (2002) 415; Phys. Rev. Lett. 89 (2002) 098701.

\bibitem{mattedi_etal} A.P. Mattedi, F.M. Ramos, R.R. Rosa and R.N. Mantegna, Physica A 344 (2004) 554.
  
\bibitem{mina_xiao} J. Mina and J.Y. Xiao, Return to RiskMetrics. The Evolution of a Standard, 
  RiskMetrics Group, New York, 2001.
  
\bibitem{nr} W.H. Press, B.P. Flannery, S.A. Teukolsky and W.T. Vetterling, 
  Numerical Recipes -  The Art of Scientific Computing, Cambridge University
  Press, New York, 1989. 
  
\bibitem{mandelbrot} B. Mandelbrot, Fractals and Scaling in Finance, Springer-Verlag, New York, 1997.

\bibitem{mantegna_prl} R.N. Mantegna and H.E. Stanley, Phys. Rev. Lett. 73 (1994) 2946. 
  
\bibitem{yahoo} \verb|http://finance.yahoo.com|
  
\bibitem{minuit} F. James, Minuit Reference Manual, CERN Program Library, 1998.

\bibitem{frey} A.J. McNeil and R. Frey, Journal of Empirical Finance 7 (2000) 271.
  
\bibitem{epjb} M.L. Goldstein, S.A. Morris and G.G. Yen, Eur. Phys. J. B 41 (2004) 255.
  
\bibitem{clementi} F. Clementi, T. Di Matteo, M. Gallegati, The power-law tail exponent of income 
  distributions, physics/0603061.
  
\bibitem{heston} S.L. Heston, Rev. Financ. Stud. 6 (1993) 327. 
  
\bibitem{borland_bouchaud} L. Borland and J. P. Bouchaud, On a multi-timescale statistical feedback 
  model for volatility fluctuations, physics/0507073
  
\bibitem{bacry_delour_muzy} E. Bacry, J. Delour and J. F. Muzy, Phys. Rev. E 64 (2001) 026103. 
  
\bibitem{pafka_kondor} S. Pafka and I. Kondor, Physica A 299 (2001) 305.
  
\bibitem{nelson} D.B. Nelson, Journal of Econometrics 52 (1992) 61.
  
\bibitem{efron_tibshirani} B. Efron and R. Tibshirani, An Introduction to the Bootstrap, Chapman \& Hall, 1993.

\end{thebibliography}
\end{document}